# Participation rates of in-class vs. online administration of low-stakes research-based assessments

Manher Jariwala[1], Jayson Nissen[2], Xochith Herrera[2], Eleanor W. Close[3], and Ben Van Dusen[2]
[1]Department of Physics, Boston University, 590 Commonwealth Avenue, Boston, MA, 02215, USA
[2]Department of Science Education, California State University Chico, 101 Holt Hall, Chico, CA, 95929, USA
[3]Department of Physics, Texas State University, 601 University Drive, San Marcos, TX, 78666, USA

This study investigates differences in student participation rates between in-class and online administrations of research-based assessments. A sample of 1,310 students from 25 sections of 3 different introductory physics courses over two semesters were instructed to complete the CLASS attitudinal survey and the concept inventory relevant to their course, either the FCI or the CSEM. Each student was randomly assigned to take one of the surveys in class and the other survey online at home using the Learning About STEM Student Outcomes (LASSO) platform. Results indicate large variations in participation rates across both test conditions (online and in class). A hierarchical generalized linear model (HGLM) of the student data utilizing logistic regression indicates that student grades in the course and faculty assessment administration practices were both significant predictors of student participation. When the recommended online assessments administration practices were implemented, participation rates were similar across test conditions. Implications for student and course assessment methodologies will be discussed.

## I. INTRODUCTION

Research-based assessments (RBAs), such as the Force Concept Inventory (FCI), the Conceptual Survey of Electricity and Magnetism (CSEM), and the Colorado Learning Attitudes about Science Survey (CLASS), are designed to measure students' knowledge of concepts or attitudes that are core to a discipline. They have been used to develop and disseminate research-based teaching practices. The demonstrated efficacy of RBAs in the research literature has led to them becoming a common method for instructors to assess student outcomes. However, many physics faculty do not use them in their classes. Faculty report various factors as barriers to using RBAs, including a lack of support in choosing assessments, guidance in administering and scoring assessments, and resources for interpreting the assessment results [1].

To address faculties' needs, educators and researchers have developed several online resources [2]. Notably, PhysPort [3] provides instructors with extensive information and guidance about selecting and administering useful assessments, and its DataExplorer tool aids instructors in analyzing and interpreting their assessment results. Similarly, in an effort to increase the use of RBAs and adoption of research-based teaching methods, the Learning Assistant (LA) Alliance, an international network of LA-using institutions [4], created the Learning About STEM Student Outcomes (LASSO) platform. LASSO is a free online platform for administering, scoring, analyzing, and tracking students' RBA scores. Administering the RBAs online removes the need to take up class time administering the assessments and may make it more attractive to more instructors to use RBAs. However, it is necessary to establish that the LASSO system's administration of computer-based tests (CBT) online outside of class provides equivalent data to that collected with paper and pencil tests (PPT) administered in class.

The research presented in this paper is part of an ongoing project designed to investigate differences in student participation and performance between CBT- and PPT-administered low-stakes RBAs. The purpose of this paper is to examine whether participation rates differ between the two different modes of administration and to identify recommended practices for instructors to maximize participation in CBT-administered assessments. We address the question of difference in performance between PPT- and CBT-administered assessments in a separate paper [5].

Most of the prior work on CBT and PPT administration focuses on K-12 classrooms administering high-stakes (graded for performance) tests in class using computers. Meta-analyses of these studies indicate that there are no systematic differences between CBTs and PPTs [6]. However, the in-class high-stakes administration in these studies made participation rates a moot point.

Similar to our study, Bonham [7] examined differences in performance on low-stakes CBT- and PPT-administered tests and attitudinal surveys, but in a college astronomy course. Like many other studies that utilize RBAs, that study did not discuss participation rates or provide total course enrollments from which participation rates could be inferred. However, if participation is related to performance, then the lower the participation rate, the more selective the sample and the less representative it is. Thus, understanding and motivating participation is key to high-quality data collection for both instructors assessing their courses and researchers pursuing scientific investigations.

## II. RESEARCH QUESTIONS

Previous work has shown that overall participation rates for low-stakes, research-based assessments given online can

be significantly lower than those given on paper [8]. In this study, we investigate the following questions:
(1) How do instructor administration practices impact participation rates for low-stakes RBAs, if at all?
(2) How are student course grades related to participation rates for low-stakes RBAs, if at all?

## III. METHODS

The study was conducted at a large regional public university in the United States. The data were collected in three different introductory physics courses: algebra-based mechanics, calculus-based mechanics, and calculus-based electricity & magnetism (E&M). A total of 25 sections across two semesters were included in the study. Algebra-based mechanics sections were traditionally taught, without clickers or required attendance. The calculus-based courses were LA-supported and used research-based instructional methods; incentives for attendance varied by instructor.

The study used a between-groups experimental design (Figure 1). Stratified random sampling created two groups within each section with similar representations across student gender, race/ethnicity, and honors status. One group completed a concept inventory online outside of class using the LASSO platform and an attitudinal survey in class using paper and pencil. The other sample completed the concept inventory in class and the attitudinal survey online outside of class. Within each course both groups completed the in-class assessment during the same class period and had the same deadline to complete the online assessments. Both conditions were repeated at the beginning and end of the semester. Paper and pencil assessments were collected by the instructors, scanned using automated equipment, and uploaded to the LASSO platform. Student assessment data was downloaded from the LASSO platform and combined with student grades and demographic data provided by the university. The data analysis did not include students who joined the class late, dropped, or withdrew. With these filters applied, the total sample was 1,310 students in 25 course sections. Of these, only 68 were students in both the calculus-based mechanics and E&M classes, and were considered as different students in our analysis.

During the first semester of data collection [8], the research team provided the instructors with little guidance in how to motivate students to complete the online assessments. Participation rates varied greatly across instructors, e.g. from 36% to 93% with a mean of 69% on the CBT pretest and 14% to 91% with a mean of 56% on the CBT posttest. The research team asked the instructors what practices they used to motivate students, and identified four different practices that increased student participation. These recommended practices are:
1. multiple email reminders,
2. multiple in class announcements,
3. participation credit for the pretest, and
4. participation credit for the posttest.

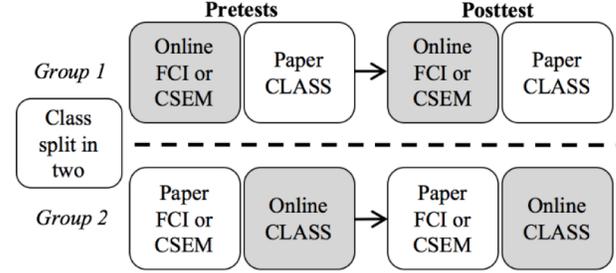

Figure 1. Design of the research conditions.

During the second semester, the research team advised all instructors to use these recommended practices to increase student participation. At the end of the semester, we interviewed the instructors to inquire what practices they had used. Analysis included both semesters of data.

In order to investigate students' participation rates in the computer versus paper and pencil assessments, we differentiated between each student's pre and post computer-based tests (CBT) and paper and pencil tests (PPT). We used the HLM 7.01 software to build hierarchical generalized linear models (HGLM) of students' participation rates for the PPT and the CBT on both the pre- and posttest. We built the HGLM as a population-averaged logistic regression model because the outcome variable was binary (i.e., whether they completed the assessment or not). In a logistic model, the coefficients for the predictors are "logits", or logarithms of the odds ratio ($P/(1-P)$):

$$\operatorname{logit}(P) = \log\left(\frac{P}{1-P}\right) \quad (1)$$

such that the probability $P$ can be calculated from the logit:

$$P = \frac{1}{1+10^{-\text{logit}}} = \frac{10^{\text{logit}}}{10^{\text{logit}}+1} \quad (2)$$

The data were nested in three levels, which are shown in Figure 1: the four measures of participation were nested within students, and the students were nested within course sections. The outcome variable for these models was whether or not students had participated in the assessment (0/1), with a separate measurement for each of the four assessments: CBT pre, PPT pre, CBT post, and PPT post. In the model, we included student's course grade as a predictor variable for all four measures and instructor practices as a predictor variable for the CBTs. The models did not include instructor practices for the PPTs because those practices were focused on improving participation on the CBTs. Instructor practices was a predictor variable at the course section level and was the cumulative number (0 to 4) of recommended practices that faculty used to motivate their students to participate in the CBTs. The models included course grades because analysis of the raw data showed that grades were positively related to participation; course grades were measured on a 0 to 4 point scale with 0 representing an F and 4 representing an A. We also looked at other student level factors like gender and race/ethnicity, but did not include them in this paper for the sake of brevity.

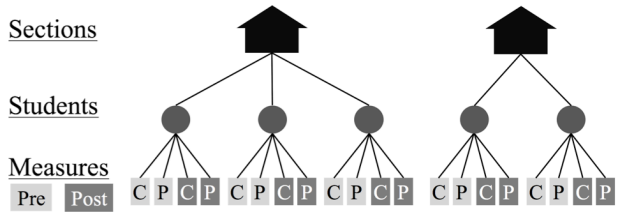

Figure 2. The data were nested in three levels. Level 1 included the CBT and PPT pre and post measures. These four measures were nested within students (level 2) who were nested within course sections (level 3).

In order to interpret the results of the model, we calculated the predicted participation rates for students by grade and instructor practices for the CBT pre- and posttests and by grade for the PPT posttest. To produce the predicted participation rates, we used the hypothesis testing function in the HLM software to generate predicted logits and standard errors for each of the combinations of variables and converted the logits to probabilities.

## IV. RESULTS

Analysis of the student data shows that the overall PPT participation rate (78.8%) is higher than the overall CBT participation rate (58.1%). Pretest participation rates were higher for both testing conditions (91.4% for PPT and 66.0% for CBT) than were posttest rates (66.2% for PPT and 50.2% for CBT). In addition, participation rates differed by course grade. With the exception of the PPT pretest (which is a proxy for attendance on the second day of class), there is a substantial range of participation rates (>25%) across grades for A to F students (Table I). While these raw results indicate that participation rates were lower on the CBTs than on PPTs, they do not account for how participation rates on the CBTs varied with instructor practices.

The results of our HGLM model of the student data, shown in Table II, indicate that the more recommended practices instructors used, the higher the participation rates were for their CBTs. Student course grades were also a statistically reliable predictor of student participation in all four conditions. As an example of the influence of instructor practices, the logit for a student who earned an "A" grade (course grade = 4) in a section where the instructor used two recommended practices for the CBT posttest would be $-1.84 + 0.36*(2) + (4)*(0.41+(2)*0.07) = 1.08$; with a logit of 1.08 equaling a probability of participating of 92%. However, for an A student in a course where no recommended practices were used, the logit would be $-1.84 + 0.36*(0) + (4)*(0.41+(0)*0.07) = -0.20$, which is equal to a probability of participating of 39%. For a C student in the same course, the logit drops even further to $-1.84 + 0.36*(0) + (2)*(0.41+(0)*0.07) = -1.02$, lowering the probability of participation down to 9%. In our model, instructor practices explained 45% of the variance.

Apart from the course section-level instructor practice variables, the student-level instructor practices variables for course grade informed the extent to which the instructors' practices differentially impacted students with different course grades. Results shown in Table 2 demonstrate that the effects were small and inconsistent (-0.04 and 0.07) and were not statistically reliable (p>0.05), indicating that using more recommended practices did not differentially motivate high and low performing students.

On their own, the size of the effect that the different coefficients have is difficult to interpret because they are expressed in logits, which use a logarithmic scale. Part of the difficulty is that the size of each coefficient cannot be directly compared because the effect of the coefficient on the probability of participating depends on the intercept. For example, a logit of 0 is a 50% probability, 1 is ~90% and 2 is ~99%. Thus, a 0.5 shift going from 0 to 0.5 (50% to 76%) is a much larger change than going from 1 to 1.5 (90% to 97%). The importance of the starting point is particularly salient for interpreting the coefficients in our HGLM model, because the intercepts for the four different measurements vary from a low of -1.84 to a high of 1.73.

Table I. Participation rates by grade, from student data.

| Grade | PPT Pre (%) | PPT Post (%) | CBT Pre (%) | CBT Post (%) |
|---|---|---|---|---|
| A | 96 | 89 | 83 | 71 |
| B | 96 | 83 | 74 | 60 |
| C | 92 | 65 | 60 | 44 |
| D | 94 | 62 | 47 | 36 |
| F | 79 | 22 | 35 | 19 |
| Drop/W | 71 | 4 | 4 | 2 |

Table II. HGLM model of student data.

| | Variables | Pretest ß | Pretest p | Posttest ß | Posttest p |
|---|---|---|---|---|---|
| CBT | For Intercept | | | | |
| | Intercept | -1.02 | 0.002 | -1.84 | <0.001 |
| | Practices | 0.29 | 0.033 | 0.36 | <0.001 |
| | For Course Grade | | | | |
| | Intercept | 0.51 | <0.001 | 0.41 | <0.001 |
| | Practices | -0.04 | 0.292 | 0.07 | 0.059 |
| PPT | For Intercept | | | | |
| | Intercept | 1.73 | <0.001 | -0.08 | 0.737 |
| | For Course Grade | | | | |
| | Intercept | 0.23 | <0.001 | 0.53 | <0.001 |

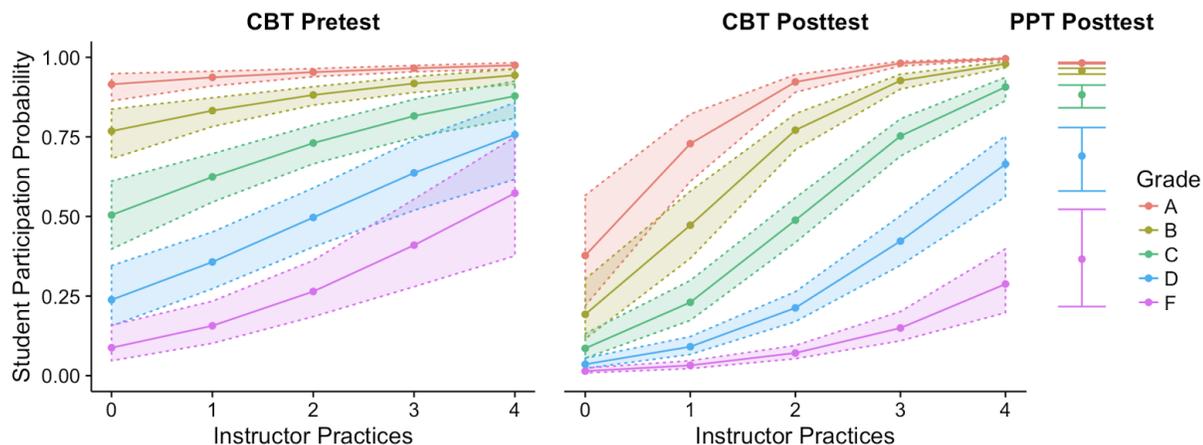

Figure 3. Predicted student participation rates on the computer-based pretest and posttest, and the paper-based posttest (far right), with 95% confidence intervals. Rates for the paper-based pretest were not plotted because the range was very small (98.0% to 99.8%).

Figure 3 illustrates the predicted student participation rate based on student course grades and the number of recommended practices that instructors used. In terms of data collection, the posttests represented the limiting case as predicted participation rates on the posttests for both the PPT and CBT were lower than on the pretests. With the exception of the PPT pretest, there was a large difference in predicted participation based on course grades. The number of recommended practices used by instructors dramatically increased predicted participation rates such that when instructors implemented all four recommended practices the participation rates of the CBT and PPT posttest were very similar. The impact of recommended instructor practices on predicted participation rates occurred for all students, but was largest for high achieving students. Relationships between student participation, grades, and instructor practices on the CBT pretest were similar to those on the CBT posttest. These results indicate that similar participation rates to those on PPT can be achieved via CBT when instructors use all four recommended practices.

## V. CONCLUSIONS

Our study shows that when faculty do not motivate student completion of online low-stakes assessments students are likely to not participate. If faculty follow a majority of our recommended practices (reminding students in class and online to participate and offering credit for participation), student participation rates for CBT matched those for PPT. This indicates that, with intention, faculty can transform their low-stakes assessments practices from in-class administrations that take up time for both students and faculty to online administrations, without lowering their student participation rates.

Our study also shows that there are meaningful discrepancies in participation rates across student grades. These differences in participation rates have implications for how student performance data is interpreted. Instructors who successfully motivate their D and F students to participate in the assessments are likely to have lower gains than their peers who only have A and B students participating. Student participation rates are rarely reported in the PER literature, much less the skewing of these rates. Ignoring the effect of this skewing could have substantial impacts on what instructors and pedagogical practices are deemed as highly effective.

The context in which our study was conducted may not be representative of physics classes at other institutions. Future research will examine the interplay of student grades and participation rates on CBT assessments at other institutions and across a range of disciplines.

## ACKNOWLEDGEMENTS

This work was funded in part by the NSF under grant nos. DUE-1525354 (MJ), DUE-1525338 (XH, JN, BVD), and DUE-808790 and DUE-1431578 (EWC). This paper is contribution No. LAA-047 of the International Learning Assistant Alliance.